\documentclass[11pt]{article}
\usepackage{mymoriond,epsfig}

\def\ra{\rightarrow}

\def\be{\begin{equation}}
\def\ee{\end{equation}}
\def\bea{\begin{eqnarray}}
\def\eea{\end{eqnarray}}

\long\def\comment#1{}

\def\la{\hbox{ \raise.35ex\rlap{$<$}\lower.6ex\hbox{$\sim$}\ }}
\def\ga{\hbox{ \raise.35ex\rlap{$>$}\lower.6ex\hbox{$\sim$}\ }}

\def\W2{{\cal W}}

\newcommand{\wjma}[6]{\left(
			   \begin{array}{ccc}
	 #1 & #2  & #3  \\
         #4 & #5  & #6
		           \end{array}
                   \right)}

\newcommand{\coefdd}[9]{{\rm d}^{\tiny\begin{array}{ccc}
                        \! #7 \! & \! #8 \! & #9
                        \end{array}}_{ \left\{\!\! \tiny\begin{array}{ccc}
         \! #1 \!\! & \!\! #2 \!\!  & \!\! #3 \! \\
         \! #4 \!\! & \!\! #5 \!\!  & \!\! #6 \!
		           \end{array} \!\!\right\} }}

\def\la{\bigl\langle} \def\ra{\bigr\rangle}

\newcommand{\Ylm}[1]{Y_{\ell_{#1}}^{m_{#1}}}

\newcommand{\almn}{a_{\ell}^{m}}

\def\be{\begin{equation}}
\def\bea{\begin{eqnarray}}
\def\ee{\end{equation}}  
\def\eea{\end{eqnarray}}

\begin{document}
\vspace*{4cm}
\title{THE BEST UNBIASED ESTIMATOR FOR THE CMB ANGULAR BISPECTRUM}

\author{ Alejandro Gangui }

\address{DARC -- Observatoire de Paris--Meudon, France}

\maketitle\abstracts{ For those angular multipoles where cosmic
variance is an issue, non-Gaussianities in the Cosmic Microwave
Background (CMB) anisotropies will be hard to detect.  Here, we
construct explicitly the best unbiased estimator for the CMB angular
bispectrum.}

While the statistical properties of the CMB anisotropies are a
powerful means to discriminate amongst the possible cosmological
scenarios, actually measuring non-Gaussianity in the data is a very
difficult task\cite{waynhu}. 
The typically small signal should be compared to the
noise and the key quantity is the signal to noise ratio. The noise
creeps into the dataset through  
instrumental errors, foregrounds contamination or incomplete sky
coverage. Add to this the so-called `cosmic variance': the fact that
we only have access to one realization of the temperature anisotropies
${\Delta }({\bf e}) \equiv { \delta T / T} ({\bf e})$
whereas theoretical predictions are expressed through ensemble
averages. It can dominate the other sources of error and therefore, if
one wants to unveil non-Gaussianity, it is necessary to address the
cosmic variance problem for those quantities characterizing a possible
non-Gaussian CMB temperature anisotropy distribution.  For that one
constructs estimators by performing spatial averages on the celestial
sphere and finds the one which has the smallest possible variance.  We
here show the {\em best unbiased estimator} for the angular bispectrum
${\cal C}_{\ell_1 \ell_2 \ell_3 }$ and we display the corresponding
cosmic variance as well. 
Recall that ${\cal C}_{\ell_1 \ell_2
\ell_3 }$ contains all the information available in the three-point
correlation function, or its variants, like the skewness, collapsed
and equilateral configurations\cite{Ganetal94}.
The present analysis borrows from recent joint work
with J\'er\^ome Martin\cite{GanMar00,GanMar01}, to whom I am greatly
{\em reconnaissant}.  

Expanding the anisotropies over the microwave sky as usual 
\begin{equation}
\label{expSW}
\Delta ({\bf e})=\sum _{\ell m}
a_{\ell  }^{m}\Ylm{}({\bf e})
\end{equation}
the first three moments can be written as
\begin{equation}
\label{propa}
\bigl\langle a_{\ell  }^{m} \bigr\rangle =0, \quad 
\bigl\langle a_{\ell _1 }^{m_1}a_{\ell _2 }^{m_2*} \bigr\rangle =
{\cal C}_{\ell_1}\delta _{\ell_1\ell_2}\delta _{m_1m_2}, \quad 
\bigl\langle a_{\ell_1 }^{m_1} a_{\ell_2}^{ m_2} a_{\ell_3}^{ m_3}
\bigr\rangle = 
\left(^{\ell_1~\,\;\ell_2~\,\;\ell_3}_{m_1~m_2~m_3}\right)
{\cal C}_{\ell_1 \ell_2 \ell_3 },
\end{equation}
where $\left(^{\ell_1~\,\;\ell_2~\,\;\ell_3}_{m_1~m_2~m_3}\right)$ is
a Wigner 3$j$-symbol. The second equation ensures the isotropy of the
CMB.  The quantity $\bigl\langle a_{\ell _1 }^{m_1}a_{\ell _2 }^{m_2*}
\bigr\rangle$ is the second moment of the $\almn$'s and ${\cal
C}_\ell$ is usually called the angular spectrum.  In the third
equation, the quantity $\bigl\langle a_{\ell_1 }^{m_1} a_{\ell_2}^{
m_2} a_{\ell_3}^{ m_3} \bigr\rangle$ is the third moment while 
${\cal C}_{\ell_1 \ell_2 \ell_3 }$ is called the angular
bispectrum.  
The presence of the 3$j$-symbol guarantees that the third
moment vanishes unless $m_1+m_2+m_3=0$ and $|\ell _i-\ell_j| \le \ell
_k \le \ell _i+\ell _j$.  Moreover, invariance under spatial
inversions of the three-point function implies the additional 
rule \cite{luo94,GanMar00} 
$\ell _1 +\ell _2 +\ell _3=\mbox{even}$, in order for the third moment
not to vanish.  Finally, from this last relation and using standard
properties of the 3$j$-symbols, it follows that the angular bispectrum
is left unchanged under any arbitrary permutation of the indices
$\ell_i$.  

Let us call ${\cal E}({\cal C}_{\ell _1 \ell_2 \ell_3})$ the estimator
for the angular bispectrum ${\cal C}_{\ell _1 \ell_2 \ell_3}$. 
The most general definition reads
\begin{equation}
\label{defE3}
{\cal E}({\cal C}_{\ell _1 \ell_2 \ell_3})\equiv \int \int \int {\rm
d}\Omega _1 {\rm d}\Omega _2 {\rm d}\Omega _3 E_{\rm S}^{\ell _1 \ell_2
\ell_3}({\bf e}_1,{\bf e}_2,{\bf e}_3) \Delta({\bf e}_1)\Delta ({\bf
e}_2)\Delta({\bf e}_3).
\end{equation}
where $E_{\rm S}^{\ell _1 \ell_2 \ell_3}$ is the weight function. The
angular bispectrum is a real quantity and so is its estimator. 
Therefore, the weight function can be taken real. It is also 
symmetric under arbitrary permutations of directions ${\bf e}_i$.  
In addition, like ${\cal C}_{\ell _1 \ell_2 \ell_3}$, the weight function 
satisfies 
$E_{\rm S}^{\ell _1 \ell_2 \ell_3} = E_{\rm S}^{\ell _2 \ell_1 \ell_3}$, 
as well as for any other arbitrary permutation of the indices $\ell_i$.
The weight function can be expressed 
on the basis of the spherical harmonics as
\begin{equation}
\label{defd3}
E_{\rm S}^{\ell _1 \ell_2 \ell_3}({\bf e}_1,{\bf e}_2,{\bf e}_3)=
\sum _{\ell_1' m_1'}
\sum _{\ell_2' m_2'}
\sum _{\ell_3' m_3'}
\coefdd{\ell_1'}{\ell_2'}{\ell_3'}{m_1'}{m_2'}{m_3'}
       {\ell_1}{\ell_2}{\ell_3}   
Y_{\ell_1'}^{m_1'}({\bf e}_1)
Y_{\ell_2'}^{m_2'}({\bf e}_2)Y_{\ell_3'}^{m_3'}({\bf e}_3).
\end{equation}
The properties of the weight function imply that the coefficients $d$ 
must satisfy 
\begin{equation}
\label{propd3}
\coefdd{\ell_1'}{\ell_2'}{\ell_3'}{m_1'}{m_2'}{m_3'}
       {\ell_1}{\ell_2}{\ell_3 *}
= (-1)^{m_1'+m_2'+m_3'}
\coefdd{\ell_1'}{\ell_2'}{\ell_3'}{-m_1'}{-m_2'}{-m_3'}
       {\ell_1}{\ell_2}{\ell_3}  
\quad
, 
\quad 
\coefdd{\ell_1'}{\ell_2'}{\ell_3'}{m_1'}{m_2'}{m_3'}
{\ell_1}{\ell_2}{\ell_3}
=
\coefdd{\ell_2'}{\ell_1'}{\ell_3'}{m_2'}{m_1'}{m_3'}
{\ell_1}{\ell_2}{\ell_3},
\end{equation}
where the last relation is in fact valid for arbitrary permutations of
any two columns of the collective subindex.  
Like the weight function, $d$ is also left invariant under 
arbitrary permutations of indices $\ell_i$ (not primed).
The estimator can be
expressed in terms of the coefficients $d$ and the $a_{\ell }^m$'s
only: inserting the expansion of the weight function in the
above expression for the estimator and using standard properties of the
spherical harmonics one obtains
\begin{equation}
\label{esti3ad}
{\cal E}({\cal C}_{\ell _1 \ell_2 \ell_3})
=
\sum _{\ell_1' m_1'}
\sum _{\ell_2' m_2'}
\sum _{\ell_3' m_3'}
\coefdd{\ell_1'}{\ell_2'}{\ell_3'}{m_1'}{m_2'}{m_3'}
       {\ell_1}{\ell_2}{\ell_3 *}
a_{\ell _1'}^{m_1'}a_{\ell _2'}^{m_2'}a_{\ell _3'}^{m_3'} .
\end{equation}
In practice, CMB observational settings are devised such that both the 
monopole and the dipole are subtracted from the anisotropy maps. 
This means that the coefficients $d$ in the last equation are only 
non-vanishing for indices $\ell_i' \ge 2$ in the collective subindex.  
Moreover, the coefficients $d$ satisfy 
$\ell _1 +\ell _2 +\ell _3=\mbox{even}$.
We must now require that our general estimator given by
Eq. (\ref{esti3ad}) be unbiased, i.e. $\langle {\cal E}({\cal C}_{\ell
_1 \ell_2 \ell_3})\rangle ={\cal C}_{\ell _1 \ell_2 \ell_3}$. This
forces the coefficients $d$ to fulfill the following constraint 
\begin{equation}
\label{cons3}
\sum _{m_1'm_2'm_3'}
\coefdd{\ell_1'}{\ell_2'}{\ell_3'}{m_1'}{m_2'}{m_3'}{\ell _1}{\ell _2}{\ell _3 *}
\wjma{\ell _1'}{\ell _2'}{\ell _3'}{m_1'}{m_2'}{m_3'}
= \delta _{\rm S}^{\ell_{i} \ell_{j}'},
\end{equation}
where we have defined a symmetrized Kr\"onecker symbol 
for the $\ell$ multipole indices only, as follows 
$
\delta _{\rm S}^{\ell_{i} \ell_{j}'} \equiv \frac{1}{6} (
\delta_{\ell_1\ell_1'}\delta_{\ell_2\ell_2'}\delta_{\ell_3\ell_3'}
+\mbox{ 5 additional permutations } ). 
$
It is easy to check that the constraint equation satisfies the
conditions imposed by Eqns. (\ref{propd3}) on the coefficients $d$. 
Using the previous properties for $d$, relabelling the indices $m_1'
\leftrightarrow m_2'$ in Eq. (\ref{cons3}) and finally noting that
$\ell_1'+\ell_2'+\ell_3'= \ell_1+\ell_2+\ell_3=\mbox{even}$, which
allows us to permute any two columns of the Wigner 3$j$-symbol, one
verifies that the left hand side of the constraint is invariant under
$\ell_1'\leftrightarrow\ell_2'$.  The same applies for any pair of
$\ell$ multipole indices and this explains the presence of the
symmetrized $\delta_{\rm S}^{\ell_{i} \ell_{j}'}$ in
Eq. (\ref{cons3}).  We see from this that all coefficients $d$ that do
not satisfy $\ell_1'+\ell_2'+\ell_3'=\mbox{even}$ do not enter the
constraint.  These terms only increase the
variance (which we want to minimize) 
and as a consequence one can take them equal to zero.

We are now in a position to calculate the variance of the
estimator. Looking at Eq. (\ref{esti3ad}) we see that this requires
the computation of the sixth moment of the $\almn$'s.
After having made use of the properties of the
coefficients $d$ and rearranging the resulting 15 terms into two groups,
straightforward algebra yields
\bea
\label{var3} 
\langle \left[ {\cal E}({\cal C}_{\ell _1 \ell_2 \ell_3})\right]^2 \rangle
&=&
\sum _{\ell_1' m_1'}
\sum _{\ell_2' m_2'}
\sum _{\ell_3' m_3'}
{\cal C}_{\ell_1'}{\cal
C}_{\ell_2'}{\cal C}_{\ell_3'} 
\\
&\times &
\biggl[ 6
\coefdd{\ell_1'}{\ell_2'}{\ell_3'}{m_1'}{m_2'}{m_3'}{\ell _1}{\ell
_2}{\ell _3 *}
\coefdd{\ell_1'}{\ell_2'}{\ell_3'}{m_1'}{m_2'}{m_3'}{\ell _1}{\ell
_2}{\ell _3} 
+
9(-1)^{m_1'+m_2'}
\coefdd{\ell_1'}{\ell_1'}{\ell_3'}{m_1'}{-m_1'}{m_3'}{\ell _1}{\ell
_2}{\ell _3 *}
\coefdd{\ell_2'}{\ell_2'}{\ell_3'}{m_2'}{-m_2'}{m_3'}{\ell _1}{\ell
_2}{\ell _3} \biggr] .
\nonumber 
\eea
The square of the variance of ${\cal E}({\cal C}_{\ell _1 \ell_2 \ell_3})$ 
is given by $\sigma ^2 _{{\cal E}({\cal C}_{\ell _1 \ell_2 \ell_3})}
=
\la \left[ {\cal E}({\cal C}_{\ell _1 \ell_2 \ell_3})\right]^2 \ra -
\la        {\cal E}({\cal C}_{\ell _1 \ell_2 \ell_3})  \ra^2$ .
Since departures from Gaussianity are expected to be small (specially
on large angular scales), higher moments will be calculated in the
mildly non-Gaussian approximation.  Within this approximation we can
write $a_{\ell}^{m} = a_{\ell}^{m (0)} + \epsilon \, a_{\ell}^{m (1)}
+ {\cal O}(\epsilon^2)$ where $a_{\ell}^{m (0)}$ is a Gaussian random
variable and the expansion parameter $\epsilon$ is small. [The
`$^{(0)}$' label will be dropped out hereafter]
Now, the term $\la \left[
{\cal E}({\cal C}_{\ell _1 \ell_2 \ell_3})\right]^2 \ra$ is of order
$\epsilon^0$ whereas the lowest non-vanishing order of $\la {\cal
E}({\cal C}_{\ell _1 \ell_2 \ell_3}) \ra^2$ is $\epsilon^2$.
Therefore, the latter one will not enter the minimization procedure and 
the variance squared will be written as 
$\sigma ^2 _{{\cal E}({\cal C}_{\ell _1 \ell_2
\ell_3})} \approx \la \left[ {\cal E}({\cal C}_{\ell _1 \ell_2
\ell_3})\right]^2 \ra$. 
This does not occur for the two-point correlator\cite{GriM}; in
that case both terms contributing to the square of the variance are
of the same order in $\epsilon$ .

With a bit of effort one can see that the various contributions of the
imaginary part of the coefficients $d$ to the two terms, $6 d^* d$ and
$9 d^* d$, only increase the variance.  Since we know that a vanishing
imaginary part does satisfy the constraint Eq. (\ref{cons3}), it can
be disregarded in the sequel.  Therefore, Eq. (\ref{var3}) can then be
written solely in terms of {\em real} coefficients $d$ as follows
\[
\sigma ^2 _{{\cal E}({\cal C}_{\ell _1 \ell_2 \ell_3})} 
=
\sum _{\ell_1' m_1'}
\sum _{\ell_2' m_2'}
\sum _{\ell_3' m_3'}
{\cal C}_{\ell_1'}{\cal
C}_{\ell_2'}{\cal C}_{\ell_3'} \biggl[6
\biggl(
\coefdd{\ell_1'}{\ell_2'}{\ell_3'}{m_1'}{m_2'}{m_3'}{\ell _1}{\ell
_2}{\ell _3 }
\biggr)^2
+9(-1)^{m_1'+m_2'}
\coefdd{\ell_1'}{\ell_1'}{\ell_3'}{m_1'}{-m_1'}{m_3'}{\ell _1}{\ell
_2}{\ell _3}
\coefdd{\ell_2'}{\ell_2'}{\ell_3'}{m_2'}{-m_2'}{m_3'}{\ell _1}{\ell
_2}{\ell _3} \biggr].
\]
Our {\em next move} now is to minimize this variance with
respect to the coefficients $d$, taking into account the constraint of
Eq. (\ref{cons3})
\begin{equation}
\label{mini3B}
\delta \biggl\{\sigma ^2 _{{\cal E}({\cal C}_{\ell _1 \ell_2 \ell_3})}
+\sum _{\ell _1'\ell _2'\ell _3'}
\lambda _{\ell _1'\ell _2'\ell _3'}^{\ell _1\ell _2\ell _3}
\biggl[
\sum _{m_1'm_2'm_3'}
\coefdd{\ell_1'}{\ell_2'}{\ell_3'}{m_1'}{m_2'}{m_3'}
       {\ell _1}{\ell_2}{\ell _3} 
\wjma{\ell _1'}{\ell _2'}{\ell _3'}{m_1'}{m_2'}{m_3'} -
\delta_{\rm S}^{\ell_{i} \ell_{j}'}
\biggr]\biggr\}=0 .
\end{equation}
Performing the variation ${\rm \delta }$ having in mind that the
symmetries of the coefficients $d$ must be respected, we get
\begin{eqnarray}
\label{mini3resu}
&&12{\cal C}_{\ell _1'}{\cal C}_{\ell _2'}{\cal C}_{\ell _3'}
\coefdd{\ell_1'}{\ell_2'}{\ell_3'}{m_1'}{m_2'}{m_3'}{\ell _1}{\ell
_2}{\ell _3} + \lambda _{\ell _1'\ell _2'\ell _3'}^{\ell _1\ell _2\ell
_3} \wjma{\ell _1'}{\ell _2'}{\ell _3'}{m_1'}{m_2'}{m_3'}
\\ 
&+& 
6(-1)^{m_2'}{\cal C}_{\ell_2'}{\cal C}_{\ell_3'}
\delta_{\ell_1'\ell_2'}\delta_{m_1'-m_2'} 
\sum _{\ell m}{\cal C}_{\ell} (-1)^{m} 
\coefdd{\ell}{\ell}{\ell_3'}{\ m \ }{\ -m \ }{\ m_3' \ }
{\ell_1 }{\ell _2 }{\ell _3 } 
+
{\tiny \Big[\begin{array}{c} 1'\to 2' \\ 2'\to 3' \\ 3'\to 1' \end{array}
\Big]}
+
{\tiny \Big[\begin{array}{c} 1'\to 3' \\ 2'\to 1' \\ 3'\to 2' 
\end{array} \Big]}
=0 .
\nonumber 
\end{eqnarray}
[$\dots$] terms are shorthand for the first one on the second line.
This formula, together with Eq. (\ref{cons3}), form a set of
equations which completely determines the best unbiased estimator. 

{}From this last equation and using the constraint Eq. (\ref{cons3})
we can get the general expression for the Lagrange multipliers.  Thus,
we multiply Eq. (\ref{mini3resu}) by the appropriate 3$j$-symbol and
we sum over the three indices $m_i'$. The first term is exactly the
constraint and produces a $\delta_{\rm S}^{\ell_{i} \ell_{j}'}$.
Using the fact that a triple sum over the $m_i$'s of the squared of a
3$j$-symbol gives unity, the second term yields the Lagrange
multipliers themselves. Unfortunately, I don't have enough space to
show that the last three terms vanish. Then, the Lagrange multipliers
are given by
\begin{equation}
\label{LagMul3}
\lambda _{\ell _1'\ell _2'\ell _3'}^{\ell _1\ell _2\ell _3}=
-12{\cal C}_{\ell _1'}{\cal C}_{\ell _2'}{\cal C}_{\ell _3'}
\delta_{\rm S}^{\ell_{i} \ell_{j}'} .
\end{equation}
Plugging this into Eq. (\ref{mini3resu}), one has
\begin{eqnarray}
\label{mini3resubis}
&&12{\cal C}_{\ell _1'}{\cal C}_{\ell _2'}{\cal C}_{\ell _3'}
\biggl[
\coefdd{\ell_1'}{\ell_2'}{\ell_3'}{m_1'}{m_2'}{m_3'}{\ell 
_1}{\ell_2}{\ell _3} 
-
\delta_{\rm S}^{\ell_{i} \ell_{j}'}
\wjma{\ell _1'}{\ell _2'}{\ell _3'}{m_1'}{m_2'}{m_3'}
\biggr]
\\ 
&+& 
6(-1)^{m_2'}{\cal C}_{\ell_2'}{\cal C}_{\ell_3'}
\delta_{\ell_1'\ell_2'}\delta_{m_1'-m_2'} 
\sum _{\ell m}{\cal C}_{\ell} (-1)^{m} 
\coefdd{\ell}{\ell}{\ell_3'}{\ m \ }{\ -m \ }{\ m_3' \ }
{\ell_1 }{\ell _2 }{\ell _3 } 
+
{\tiny \Big[\begin{array}{c} 1'\to 2' \\ 2'\to 3' \\ 3'\to 1' \end{array}
\Big]}
+
{\tiny \Big[\begin{array}{c} 1'\to 3' \\ 2'\to 1' \\ 3'\to 2' \end{array} \Big]}=0 .
\nonumber 
\eea
This is the final equation to be solved in order to determine the 
best unbiased estimator. A solution is 
\begin{equation}
\label{sold3}
\coefdd{\ell_1'}{\ell_2'}{\ell_3'}{m_1'}{m_2'}{m_3'}{\ell_1 }{\ell_2 }{\ell_3 }
=\wjma{\ell _1'}{\ell _2'}{\ell _3'}{m_1'}{m_2'}{m_3'}
\delta_{\rm S}^{\ell_{i} \ell_{j}'} ,
\end{equation}
which leads to
\begin{equation}
\label{solest3}
{\cal E}_{\rm Best}({\cal C}_{\ell_1 \ell_2 \ell_3 })
= \sum _{m_1' m_2' m_3'}
\wjma{\ell_1 }{\ell_2 }{\ell_3 }{m_1'}{m_2'}{m_3'}
a_{\ell_1 }^{m_1'}a_{\ell_2 }^{m_2'}a_{\ell_3 }^{m_3'} .
\end{equation}
Seems familiar? An estimator restricted to the diagonal case $\ell
_1=\ell _2=\ell _3$ (and then extended to $\ell_2 = \ell_1+2$ and
$\ell_3 = \ell_1-2$) has been proposed \cite{Feretal98,newturn} for
${\cal B}_{\ell }\equiv {\cal C}_{\ell \ell \ell}$. ~\footnote{${\cal
B}$ like ${\cal B}$ispectrum (like spe${\cal C}$trum) $\ldots$  
or ${\cal C}$, ${\cal B}$, ${\cal
A}$?, $\ldots$ .  Hope the ${\cal T}$rispectrum will be called ${\cal
T}$. \\ 
Still, since no ambiguity arises I stick to 
${\cal C}_{\ell_1 \ell_2 \ell_3 }$ (for now).}
The aim of these authors was not to seek the best estimator, but to
use the corresponding, say, ${\cal E}({\cal B}_{\ell })$ to analyse
the non-Gaussian features of the 4-yr COBE-DMR 
data (see also\cite{Hea98,hobson98,BroTeg99,BZG}).  
While their
estimator is {\em not} unbiased, for it does not satisfy the constraint
(\ref{cons3}), by just removing an overall prefactor one gets our best
unbiased estimator ${\cal E}_{\rm Best}({\cal C}_{\ell_1 \ell_2 \ell_3
})$, Eq. (\ref{solest3}).

We now know the best unbiased estimator for ${\cal C}_{\ell_1 \ell_2
\ell_3 }$ and then we can compute its variance, the smallest one
amongst all possible estimator variances, which yields
\be
\label{vari2gm}
\sigma ^2 _{{\cal E}_{\rm Best}({\cal C}_{\ell _1 \ell_2 \ell_3})}
=
{\cal C}_{\ell_1} {\cal C}_{\ell_2} {\cal C}_{\ell_3}
(1+\delta_{\ell_1\ell_2}+\delta_{\ell_2\ell_3}+\delta_{\ell_3\ell_1}
+ 2 ~ \delta_{\ell_1\ell_2}\delta_{\ell_2\ell_3}) .
\ee
We like to dub this (the square of) the 
`bispectrum cosmic variance' in perfect analogy with
$\sigma^2_{{\cal E}_{\rm Best}({\cal C}_{\ell})}
= 2 {\cal C}_{\ell}^2 / (2\ell+1)$, which is (the square of) the 
variance of the best unbiased estimator for the angular spectrum,
commonly known as the `cosmic variance'. 

\noindent
{\em Acknowledgments}

\noindent 
I'd like to thank my collaborator J. Martin for extensive discussions
and A. Heavens for useful correspondence.
This work was partially financed with funds from the World Lab.

\end{document}